\documentclass[aps,twocolumn,epsfig
]{revtex4}
\usepackage{graphicx}
\usepackage{amsmath}
\usepackage{latexsym}
\usepackage{amsfonts}
\usepackage{amssymb}
\usepackage{array}
\usepackage{color}
\usepackage{subfigure}
\usepackage{verbatim}
\newcommand{\ket}[1]{\left\vert#1\right\rangle}
\newcommand{\bra}[1]{\left\langle#1\right\vert}

\newcommand{\braket}[2]{\langle#1|#2\rangle}

\begin{document}
\newcommand{\Q}[1]{{\color{red}#1}}
\newcommand{\blue}[1]{{\color{blue}#1}}
\newcommand{\red}[1]{{\color{red}#1}}
\newcommand{\green}[1]{{\color{green}#1}}
\newcommand{\update}[1]{{\color{magenta}#1}}
\newcommand{\Change}[1]{{\color{green}#1}}
\title{
Nonlinear potential of quantum oscillator induced by single photons}

\author{Kimin Park}
\email{park@optics.upol.cz}
\author{Petr Marek}
\author{Radim Filip}
\affiliation{Department of Optics, Palack\'y University, 17. listopadu 1192/12, 77146 Olomouc, Czech Republic}
\date{\today}

\begin{abstract}
Experimental investigation of the nonlinear dynamics of a quantum oscillator is a long standing goal of quantum physics. We propose a conditional method for inducing an arbitrary nonlinear potential on a quantum oscillator weakly interacting with light. Such an arbitrary nonlinear potential can be implemented by sequential repetition of an elementary conditional $X$-gate.
To implement the $X$-gate, a single photon is linearly coupled to the oscillator and is subsequently detected by optical homodyne detection.
\end{abstract}

\maketitle

\section{Introduction}
In quantum physics it is crucial to be able to precisely manipulate quantum systems. This ability is the key both to experimental tests of fundamental natural principles and to the actual development of quantum technology. The ultimate aim in this direction is the implementation of a variety of nonlinear transformations. One way of approaching this daunting task lies in disassembling general operations into elementary building blocks. For two-level (qubit) quantum systems, such building blocks are the single qubit rotations and the two-qubit controlled NOT operation \cite{Nielsen}. In a similar vein, the basic building blocks for continuous variable harmonic oscillator systems \cite{CV1,CV2} are the operations imposing quadratic and cubic potentials \cite{Gottesman2001,Bartlett2002}. The quadratic potential inducing Gaussian operations can be considered readily available. A general method of achieving any form of quadratic potential uses squeezed states of light which interact with the oscillator and are subsequently measured by an optical homodyne detection \cite{Filip2005,Yoshikawa2007,Yoshikawa2008,Yoshikawa2011}.

However, squeezed states of light and optical homodyne detections are not sufficient resources to induce highly nonlinear potentials, such as the cubic one. Since fully deterministic implementation of cubic nonlinearity is a very challenging task \cite{Marek2011}, it is important to be able to induce a non-linear potential on a quantum oscillator at least conditionally, as it is currently the only feasible way for studying the nonlinear quantum dynamics.
A straightforward, but complicated way is to use the typical decomposition of quantum operations relying on annihilation $\hat{a}$ and creation $a^{\dagger}$ operators \cite{Dakna1997,Dakna1998,Dakna1999,Ourjoumtsev2006,Nielsen2006,Parigi2007}. These operators with clear Fock state interpretation play an important role in phase insensitive applications \cite{Fiurasek2009}, such as entanglement distillation \cite{Opatrny2000,Takahashi2010} or a version of the noiseless amplification \cite{Marek2010,Usuga2010,Zavatta2011}.

In this article we present a complementary approach which allows inducing an arbitrary nonlinear potential $V(\hat{X})$ on a quantum oscillator by sequential application of the position operator $\hat{X}=(\hat{a}+{\hat{a}}^{\dagger})/\sqrt{2}$ which was also denoted as the orthogonalizer~\cite{Vanner2013a}, by an operation which we will call as the X-gate. An optical scheme  to achieve an operation $m^*\hat{a}+n^*\hat{a}^\dagger$ was also proposed in \cite{SLeePRA2010}
 using a standard approach with nonlinear resources, while our scheme is more compact and suitable for sequential application. The main benefit of using X-gate instead of the annihilation or creation operators is that the former can be naturally extended to physical systems other than light, such as mechanical oscillators or clouds of atoms, and that the exact form of the potential can be adjusted at will. As the resource for the X-gate we are going to use single photon guns \cite{Kuhn2002,Legero2004,Chen2006,Hijlkema2007,Bozyigit2011,Fortsch2013,Collins2013,He2013},
which were recently extensively developed for broad class of applications. We analyze the performance and feasibility of this methodology with regard to realistic experimental tools and emphasize two exemplary applications: generation of the cubic nonlinearity and efficient state preparation of non-Gaussian states.

In Sec. II, we analyze how to implement the X-gate in various ways. We investigate the performance of our gate in realistic situations in Sec. III. Applications of our gates are summarized in IV. In Sec. V we conclude.

\section{Implementation of X-gate}
\subsection{Oscillator in a nonlinear potential}
The quantum oscillator with a Hamiltonian operator $\hat{H}=\hbar\omega\left(\hat{a}^{\dagger}\hat{a}+\frac{1}{2}\right)+V(\hat{X})$, where $\hat{X}$ is the position operator and $V(\hat{X})$ is a nonlinear potential, contains a mixture of free linear evolution with frequency $\omega$ and nonlinear dynamics induced by $V(\hat{X})$. To obtain the pure effect of a nonlinear potential on a quantum system, we assume the limit $\omega\rightarrow 0$ of low-frequency oscillator  evolving very slowly.
In this limit, the unitary evolution operator $U(\hat{X},\tau)=e^{-\frac{i}{\hbar}V(\hat{X})\tau}$, where $\tau$ is the time duration of evolution in the potential, preserves the statistics of position and affects only the statistics of the complementary variable described by the momentum operator $\hat{P}=(\hat{a}-\hat{a}^\dagger)/\sqrt{2}i$.

The evolution operator can be approximated by a Taylor series { $U(\hat{X},\tau)=\sum_{k=0}^{\infty}\frac{U^{(k)}(\bar{X})}{k!}(\hat{X}-\bar{X})^k$ }around the initial mean position $\bar{X}$ of the oscillator. The finite truncation of this Taylor series can be expanded as $U(\hat{X},\tau) = \prod_{k=0}^{N}(1+\lambda_k \hat{X})$ using the general theorem of algebra, where $\lambda_k$'s are related to the complex roots of the polynomial, $U(-\lambda_k^{-1},\tau) = 0$. Any dynamics imposed purely by the nonlinear potential can therefore be decomposed to a sequence of the {\em non-unitary X-gates} $\hat{\mathcal{A}}_X(\lambda_k)=1+\lambda_k \hat{X}$ controlled by {\em complex} parameters $\lambda_k$'s. For a purely imaginary $\lambda_k$ with the magnitude close to zero, the operation $\hat{\mathcal{A}}_X(\lambda_k)$ is close to a unitary displacement operator. For a larger magnitude of purely imaginary or real $\lambda_k$, however, the X-gate is inherently probabilistic and its action is non-trivial. Our approach suggests how to implement the individual X-gates which are applied sequentially with variable complex parameters $\lambda_k$ to mimic the behavior of slowly evolving quantum oscillators in the nonlinear potential. 

\subsection{Coupling an oscillator to light}
 Implementation of an individual X-gate exploits one of two kinds of coupling between a quantum oscillator and a single mode of electromagnetic radiation. Under the approximation of weak coupling for which the time duration is short enough, the interaction can be represented by a unitary operator derived from one of two possible interaction Hamiltonians. The beam splitter (BS) interaction with $\hat{H}_\mathrm{BS}=i\kappa_\mathrm{BS} (\hat{a}^{\dagger}\hat{b}_L-\hat{b}_L^{\dagger}\hat{a})$, where $\hat{a}$ is annihilation operator of the quantum oscillator and $\hat{b}_L$ is annihilation operator of the single mode L of radiation, represents a natural coupling between different modes of radiation varying in polarization, spatial properties, or frequency \cite{Zaske2012}. It can be also used to describe coupling with continuous-wave or semi-continuous-wave regime of mechanical oscillator \cite{Hofer,Verhagen2012}. The second kind of coupling is the quantum non-demolition (QND) coupling given by $\hat{H}_\mathrm{QND}=i\kappa_\mathrm{QND} (\hat{a}^{\dagger}+\hat{a})({\hat{b}_L}^{\dagger}-\hat{b}_L)/2$. This type of interaction naturally appears for the coupling with spin ensembles \cite{Hammerer2010,Sewell2013} and  the pulsed regime of mechanical oscillators \cite{Vanner2011,Vanner2013a,Vanner2013b}.


\subsection{Elementary X-gate based on BS coupling}
\begin{figure}[th]
\includegraphics[width=250px]{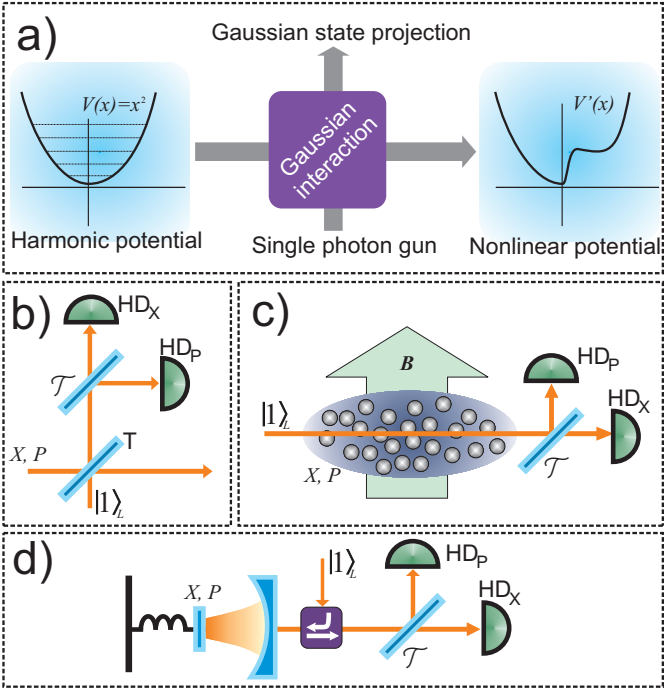}
\caption{(Color online) a) Concept of the implementation of a nonlinear potential by single photon guns; b) X-gate for a single mode of light using BS-type coupling; c) X-gate for a collective spin of cloud of atoms in magnetic field; d) X-gate for vibration mode of a mechanical oscillator. }
\label{fig:scheme}
\end{figure}

We shall start by explaining the implementation of the X-gate for the BS coupling, because it plays prominent role in all-optical implementations, which are in turn a natural platform for experimental tests of the method. For reasons which will become clear later, we generalize the X-gate $\hat{\mathcal{A}}_X(\lambda)=1+\lambda \hat{X}_\theta$  to a more general class of operations:
\begin{equation}\label{operation}
    \hat{\mathcal{A}}(\lambda_-,\lambda_+) = 1+\lambda_-\hat{a}+\lambda_+\hat{a}^\dagger,
\end{equation}
where $\hat{a}$ and $\hat{a}^{\dag}$ are the annihilation and creation operators, respectively. Here $\lambda_+$ and $\lambda_-$ are complex numbers which can be adjusted at will. The conceptual scheme for implementing the ideal operation (\ref{operation}) is depicted in Fig.~\ref{fig:scheme}. This scheme is a measurement-induced operation which is composed of the main implementation step and the correction step. In the first step, the input oscillator mode interacts with the ancillary mode L in the single photon state $|1\rangle_L$. The ancillary mode L is subsequently measured by a setup which contains beam splitters and homodyne detectors, and the state of the oscillator mode is post-selected when specific values are detected.  This process can be expressed as the projection onto a Gaussian state $\ket{\zeta}$, which is represented by an operator $~_L\langle \zeta | U_{BS}|1\rangle_L$. Here $\hat{U}_\mathrm{BS}=\exp(-i \hat{H}_\mathrm{BS}t) = T^{\hat{n}}e^{-R^* \hat{b}_L^\dagger \hat{a}}e^{R \hat{b}_L \hat{a}^\dagger}T^{-\hat{n}_L}$ stands for the unitary operator of the beam splitter with transmission coefficient $T=\cos\kappa t_\mathrm{BS}$ which is coupling the ancillary mode to the oscillator. Here $\hat{n} = \hat{a}^{\dag}\hat{a}$ and $\hat{n}_L = \hat{b}_L^{\dag}\hat{b}_L$.

The projection $|\zeta\rangle_L\langle \zeta |$ can be implemented by an unbalanced heterodyne detection - the ancillary mode L is split at an unbalanced beam splitter with transmission and reflection coefficients $\mathcal{T}$ and $\mathcal{R}$, and optical homodyne detections of complementary quadratures $\hat{X}_L=(\hat{b}_L +\hat{b}^{\dag}_L)/\sqrt{2}$ and $\hat{P}_L=(\hat{b}_L -\hat{b}^{\dag}_L)/\sqrt{2}i$ are performed on each output port. Such a measurement can be represented by the projection onto a state:
\begin{align}
&_L\bra{x}_{L'}\bra{p}U_\textrm{BS}\ket{0}_L'=\nonumber\\
&_L\bra{0}\exp[-\frac{x^2+p^2}{2}+\sqrt{2}(x \mathcal{T}+i p \mathcal{R}^*)b_L+\frac{\mathcal{R^*}^2-\mathcal{T}^2 }{2}b_L^{2}]\nonumber\\
&\propto _L\bra{0}\exp[A^* b_L+B^* b_L^{2}]\equiv _L\bra{A,B},
\end{align}
where $A=\sqrt{2}(x \mathcal{T}-i p \mathcal{R})$ and $B=2^{-1}(\mathcal{R}^2-\mathcal{T}^2)$ are complex measurement parameters with $-1/2<|B|<1/2$, whose phases $\arg{A}$ and $\arg{B}$ can be chosen arbitrarily.

The full operation by the homodyne detection looks like:
\begin{eqnarray}
\label{eq:scheme}
_L\langle x_\theta|U_\mathrm{BS}\ket{1}_L\propto&T^{\hat{n}}\exp[-\sqrt{2}x_\theta e^{-i\theta} R^* \hat{a}-\frac{R^{*2}e^{-2i\theta} \hat{a}^2}{2}]\nonumber\\
&(\frac{\sqrt{2}}{T}x_\theta+\frac{R^*e^{-i\theta}}{T}\hat{a}+\frac{R e^{i\theta}}{T} \hat{a}^\dagger), 
\end{eqnarray}
and the complete operation by the heterodyne measurement is  summarized as:
\begin{eqnarray}\label{eq:scheme2}
_L\bra{A,B}U_\mathrm{BS}\ket{1}_L=
\exp[A^* \frac{R}{T} \hat{a}+B^*\frac{R^2}{T^2} \hat{a}^2]\times\nonumber\\
T^{\hat{n}-1}(A^*+2B^* R^* \hat{a}+R \hat{a}^\dagger).
\end{eqnarray}
The operator (\ref{eq:scheme2}) is composed of three parts: the ideal operation $A^*+2B^* R^* \hat{a}+R \hat{a}^\dagger$ consisting of the proper superposition of annihilation and creation operators, the error operator $\exp[A^* R \hat{a}+B^*R^2 \hat{a}^2]$, and another error operator $T^{\hat{n}-1} $ which we will denote as pure attenuation. These two sources of error need to be considered separately, as each of them possesses very different properties.
The error term $\exp[A^* R \hat{a}+B^*R^2 \hat{a}^2]$ can be compensated by a correction operation using optical ancilla $L'$ in the vacuum state:
\begin{equation}
_{L'}\bra{A',B'}U_\mathrm{BS}\ket{0}_{L'}=T'^{\hat{n}}\exp[A'^* R' \hat{a}+B'^*R'^2 \hat{a}^2],
\end{equation}
which is implemented in the same way as the main gate in Eq.~(\ref{eq:scheme2}), only with a replacement of single photon by the vacuum state in the ancillary mode. With $A'=-A/T$, $B'=-B/T^2$ and $R'=R$ we can erase the error and obtain an approximate version
\begin{equation}
\hat{\mathcal{A}}_\mathrm{BS}=(T'T)^{\hat{n}}(A^*+2B^* R^* \hat{a}+R \hat{a}^\dagger)
\label{eq:aftercorrection}
\end{equation}
of the generalized X-gate (\ref{operation}) using the BS coupling.
The desired gate is accompanied by an increased noiseless attenuation $(T'T)^{\hat{n}}$ as an unavoidable cost of transforming an ill-behaved error into a well behaved one. It should be noted that in the case of a highly transmissive beam splitter $R\ll 1$, all sorts of errors become less prominent even up to the point when the correction step is not necessary. The cost of this strategy is the diminished success rate and a high sensitivity to the quality of ancilla.

The noiseless attenuation error caused by $(T'T)^{\hat{n}}$ becomes significant when the elementary X-gates are combined into a more complicated function. For this purpose we have to apply relations  $T^{\hat{n}} \hat{a} = \hat{a}T^{\hat{n}-1}$ and $T^{\hat{n}} \hat{a}^{\dag} = \hat{a}^{\dag} T^{\hat{n}+1}$ to move the attenuation term. As a consequence, an arbitrary polynomial  $\prod_{i=0}^{N}(1+\lambda_i \hat{X})$ needs to be implemented as:
\begin{eqnarray}
\prod_i T_i^{\hat{n}}\left(1 + \lambda_i \frac{\mathbb{T}_i\hat{a}+\mathbb{T}_i^{-1}\hat{a}^\dagger}{\sqrt{2}}\right)=\left[\prod_i^N(1+\lambda_i \hat{X})\right] \mathbb{T}_N^{\hat{n}},
\end{eqnarray}
where $\mathbb{T}_i=\prod_{j=1}^i T_j$. As can be seen, the noiseless attenuation is effectively applied only once, solely on the initial state. In principle it can be approximatively compensated by the noiseless amplification conditionally approaching operation $G^{\hat{n}}$ with $G>1$ \cite{Ralph2009}. On the other hand, the noiseless attenuation has a very clear Fock space interpretation and it is always acting in a predictable manner. In many experiments it can be therefore taken into account and compensated by manipulating the measured data.

\subsection{Elementary X-gate based on QND coupling}
Although the QND coupling can be established between different modes of radiation \cite{Filip2005,Yoshikawa2008}, it is much more important in experiments with atomic spin ensembles \cite{Hammerer2010,Sewell2013}, or pulsed regime of mechanical oscillators \cite{Vanner2011,Vanner2013a,Vanner2013b}, where it appears naturally. Adapting the X-gate for this coupling therefore allows expanding the methods of quantum optics even to these systems. For the QND coupling, represented by the unitary operator $\hat{U}_\mathrm{QND}=e^{-i \kappa \hat{X}\hat{P}_L}$, where $\hat{X}=(\hat{a} +\hat{a}^{\dag})/\sqrt{2}$ and $\hat{P}_L=(\hat{b}_L -\hat{b}^{\dag}_L)/\sqrt{2}i$, of optical mode $L$ to the oscillator the complete gate can again be expressed as:
\begin{eqnarray}
_L\bra{A,B}U_\mathrm{QND} \ket{1}_L\propto
\exp[\frac{A \kappa}{\sqrt{2}} \hat{X}+(\frac{B}{2}-\frac{1}{4}) \kappa^2 \hat{X}^2]\times\nonumber\\
\left\{A+\kappa (2B-\frac{1}{\sqrt{2}})\hat{X}\right\},
\end{eqnarray}
where $A$ and $B$ are the same as before and $\kappa=\kappa_\mathrm{QND}t$. In a similar manner as for the BS interaction, the correction operation required to eliminate the error term $\exp[\frac{A \kappa}{\sqrt{2}} \hat{X}+(\frac{B}{2}-\frac{1}{4}) \kappa^2 \hat{X}^2]$ is $~_L\bra{-A,-B}\hat{U}_{QND} \ket{0}_L=\exp[-A \frac{\kappa}{\sqrt{2}} \hat{X}-(B/2+1/4) \kappa^2 \hat{X}^2]$, which is implemented using another QND interaction with optical mode being in vacuum state. The redundant $\exp[-\kappa^2\hat{X}^2/4]$ can be in part compensated by squeezing the ancillary state, whose effect can be described by $\exp[\tanh r\kappa^2\hat{X}^2/4]$. In contrast to the BS type of coupling to the optical mode, after erasing the error term, we approach the ideal X-gate without the noiseless attenuation errors.
Moreover, the X-gate can be also implemented by replacing the homodyne detection by a photon number resolving detector and changing the ancilla. The resulting gate,
\begin{align}
_L\bra{0}U_\mathrm{QND}(\ket{0}_L+c_1\ket{1}_L)=\exp[-\frac{\kappa^2\hat{X}^2}{4}](1+c_1\frac{\kappa}{\sqrt{2}} X_A)
\end{align}
has always a non-zero probability of success. This approach will become fully feasible with advent of efficient photon number resolving detectors.

\section{Realistic considerations}
\subsection{Requirements on quality of single photons}
The single photons employed by the $X$ gate are an experimental resource sensitive to imperfections. They usually do not appear in the pure form $\ket{1}_L$, but rather in a mixture $\eta|1\rangle_L\langle 1| + (1-\eta)|0\rangle_L\langle 0|$ \cite{Lvovsky2001} which may reduce the quality of the gate. To quantify the quality of single photon that is necessary for successful implementation of X-gate, we compare the performance of the gate with methods using coherent state ancillas. The required quality of the single-photon gun is then characterized by the critical efficiency $\eta_c$, the value of $\eta$ for which the fidelity of the gate is equal to the classical threshold.

\subsection{Performance analysis and the classical threshold}
For the analysis of performance, we apply the X gate to a set of quantum states and compare their fidelities. For this analysis it is advantageous to consider quantum states which are orthogonalized by the X operation, because then the operation $1+\lambda X$ effectively creates a qubit, whose fidelity has a good operational meaning. The states which satisfy this criterion are the coherent states with purely imaginary amplitudes, $|\beta\rangle$ with $\beta = i|\beta|$; single photon state $|1\rangle$; and the squeezed state $|\xi\rangle =\exp[-\xi/2 \hat{a}^{2\dagger}+\xi/2 \hat{a}^2]\ket{0}$. For these states, the fidelities are compared to the classical benchmark which is obtained by considering the gate with only a classical state used as an ancilla. As any classical state can be represented as a mixture of coherent states, it is sufficient to consider a coherent state as the ancilla and maximize over its amplitude. The operation with the classical resource can be written as
\begin{align}
&_2\langle x=0| \hat{U}_\mathrm{BS}\ket{\alpha}_2\propto\nonumber\\
&T^{\hat{n}}\exp[\alpha R T^{-1} \hat{a}^\dagger]\exp[-\frac{R^2}{2}\hat{a}^2]\exp[\alpha R T \hat{a}]=\nonumber\\
&\exp[\alpha R \hat{a}^\dagger]\exp[\alpha R  \hat{a}]\exp[-\frac{R^2}{2T^2}\hat{a}^2]T^{\hat{n}}.
\end{align}
Note that  it is simply impossible to obtain the desired X operation perfectly with a classical resource regardless of any correction we may apply.

Another benchmark is obtained by trying to achieve the target operation by using only unitary Gaussian operations - displacement and squeezing. These operations are experimentally feasible, but on their own they are not sufficient for obtaining any kind of higher order nonlinearity. For the target single photon input state, the Gaussian benchmark is $0.82$, which leads to $\eta_c\approx0.7$ for $T\approx 0.734$. For other input states we are considering, these unitary Gaussian operations give a lower benchmark and need not to be considered. 

With a realistic resource single photon, the full gate (with the correction) transforms the input state $|\psi\rangle$ into 
\begin{align}
\rho &\propto T^{\hat{n}}(\eta R^2/\lambda^2 T^2 (1+\lambda\hat{a}\pm\lambda a^\dagger)\ket{\psi}\bra{\psi}(1+\lambda\hat{a}^\dagger\pm\lambda a)\nonumber\\
&+(1-\eta)\ket{\psi}\bra{\psi})T^{\hat{n}}.
\end{align}
  We notice that for a very small $T\ll 1$, the effect of lower $\eta$ in single photon generation can be completely ignored, and a perfect target operation is achieved regardless of $\eta$, however, only at the cost of a significant noiseless attenuation. This can be seen as a conditional transformation of the resource state's impurity to noiseless attenuation, which does not significantly reduce purity of the state. This is a valuable strategy if the noiseless attenuation does not play an important role. However, if this is not the case or if the attenuation cannot be very well compensated by a suitable noiseless amplification, the efficiency $\eta$ remains important.

\begin{figure}[thp]
\subfigure[~$\ket{\beta=0.1}$]{
\includegraphics[width=115px]{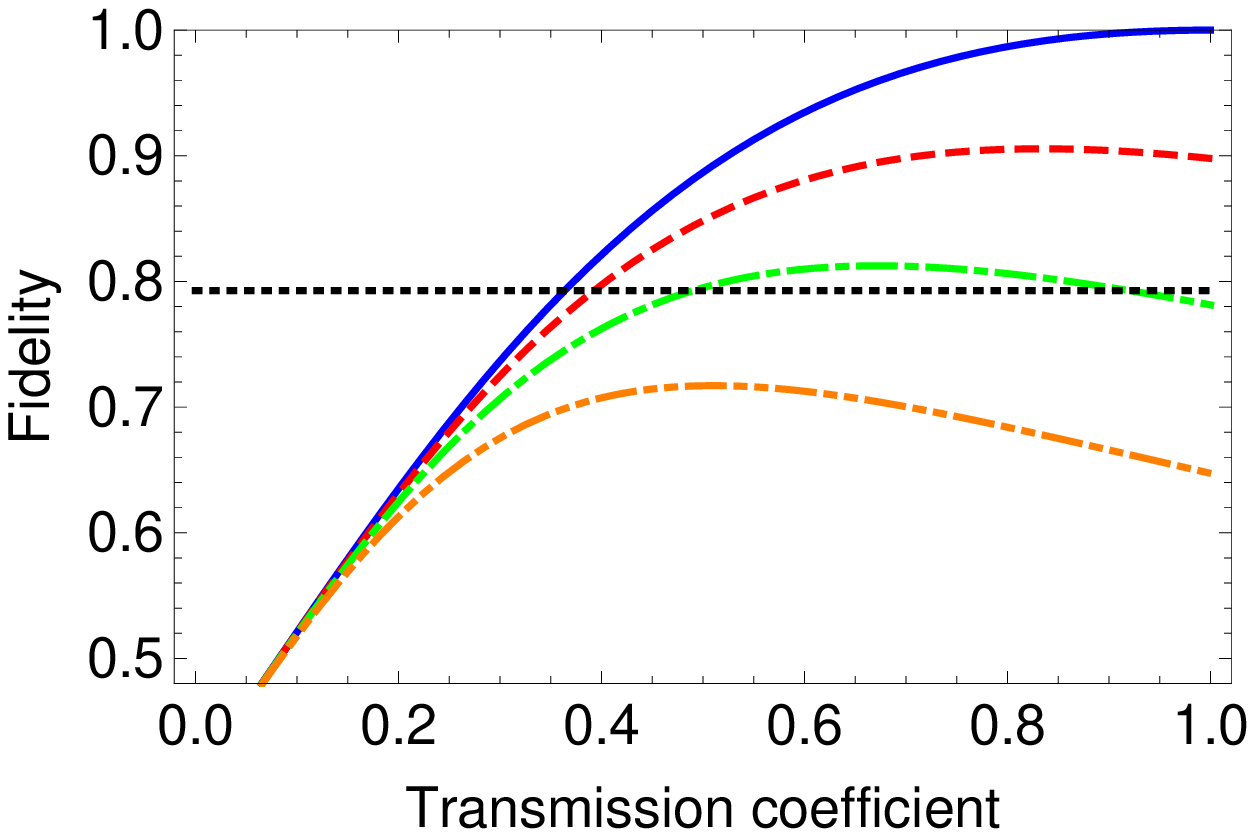}
}
\subfigure[~$\ket{\beta=1}$]{
\includegraphics[width=115px]{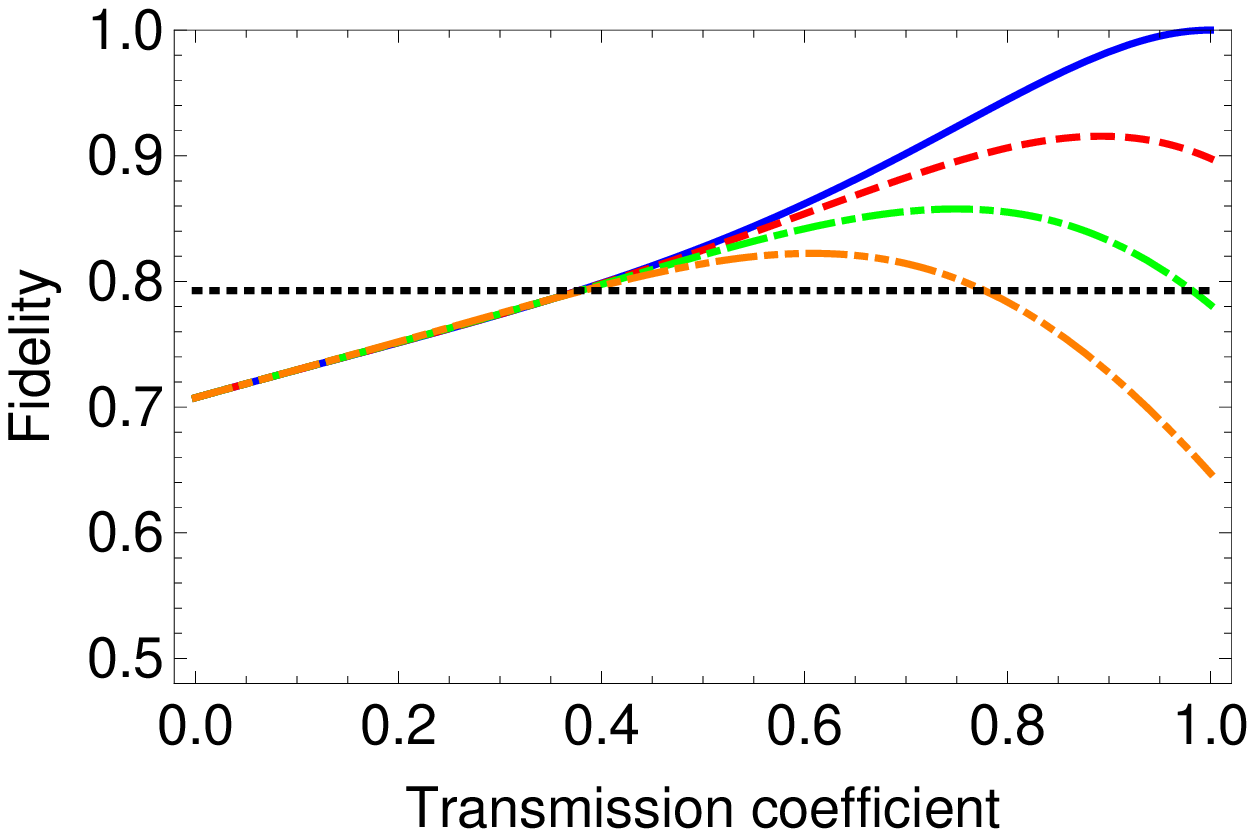}
}
\subfigure[~$\ket{\xi=0.1}$]{
\includegraphics[width=115px]{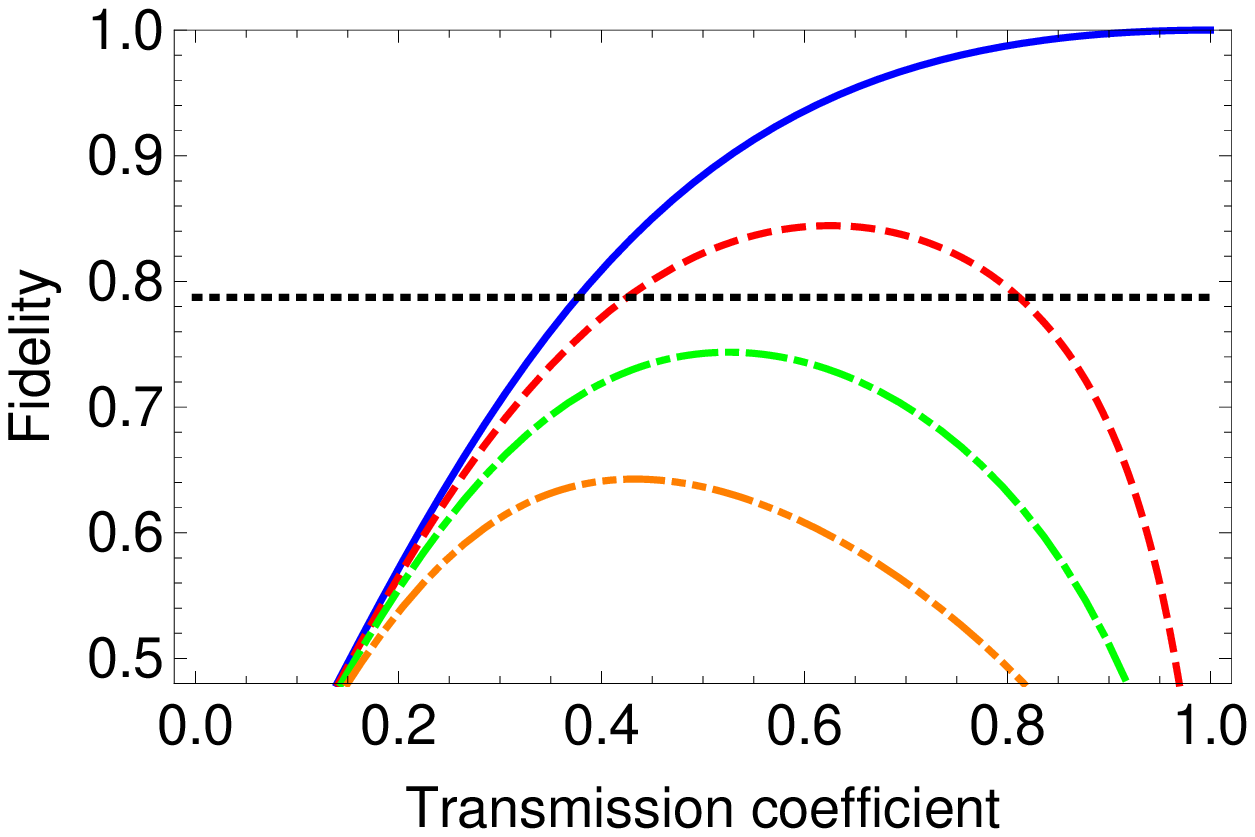}
}
\subfigure[~$\ket{1}$]{
\includegraphics[width=115px]{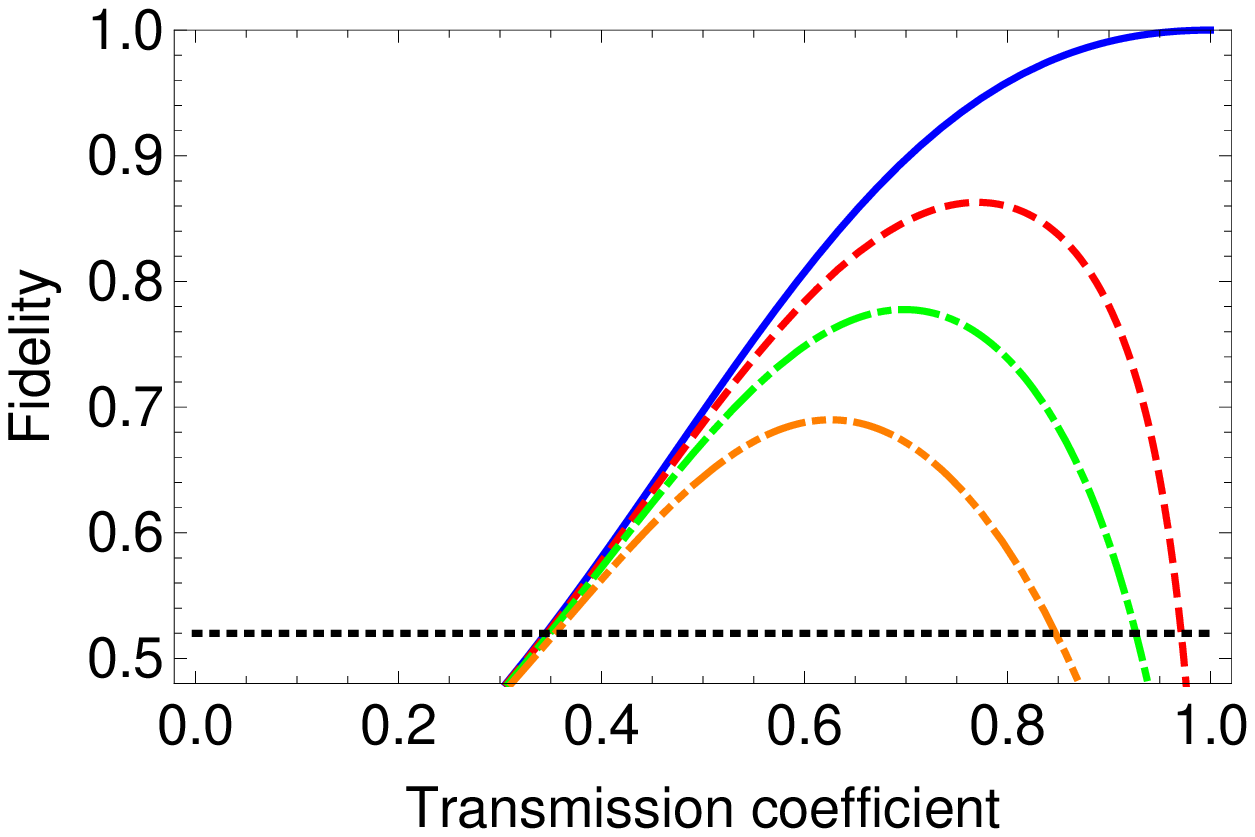}
}
\caption{
Fidelity vs transmission coefficient $T$ for operation $1+1.5\hat{a}-1.5\hat{a}^\dagger$ on coherent state inputs (a) $\ket{\beta=0.1}$ and (b) $\ket{\beta=1}$, squeezed state inputs (c) $\ket{\xi=0.1}$, and (d) single photon $\ket{1}$, with imperfect single photon ancilla $\eta\ket{1}_L\bra{1}+(1-\eta)\ket{0}_L\bra{0}$. Near $T\approx 1$, the fidelity is high for $\eta=1$ (blue), but drops rapidly when the ancilla is imperfect ($\eta=0.8$ (red), $0.6$ (green) and $0.4$ (orange)) below the classical benchmark (dotted). The values for classical benchmark are $0.79$ for coherent states and squeezed states, and $0.52$ for single photon state.}
\label{fig:etac}
\end{figure}

In Fig.~\ref{fig:etac} we show the analysis of a trial gate operation $1+\lambda\hat{a}-\lambda\hat{a}^\dagger$ applied to selected quantum states for various levels of quality of the single photon ancilla, where their fidelities with the ideal states are compared to the classical threshold.  When $\lambda$ is as small as $0.1$, the operation is generally well simulated by a displacement operator, and the classical threshold fidelity is typically as high as $0.99$. For large $\lambda=1.5$ on the other hand,  $\eta_c\approx0.55$ for a coherent state $\ket{\beta=0.1}$, and $\eta_c\approx0.35$ for a coherent state $\ket{\beta=1}$.  For a single-mode squeezed vacuum state input $\ket{\xi}=\hat{S}(\xi)\ket{0}$,   $\eta_c\approx0.7$ for $\ket{\xi=0.1}$, and $\eta_c\approx0.6$ for $\ket{\xi=1}$. For single photon input $\ket{1}$, for $T\approx 0.45$ we can achieve $\eta_c=0.12$. Therefore, with current quality of single-photon gun our scheme can surpass classical resources rather easily. It is therefore feasible to experimentally observe the non-classical performance of elementary X-gate with limited $|\lambda|$. Note that the performance of the gate for large $\lambda$ can be used as a very strict operational measure of single photon states, as in this case even resource states with significant negativity in Wigner function \cite{Lvovsky2001} might not be sufficient for beating the classical threshold.

\subsection{Success Rate vs. Fidelity}
\begin{figure}[tp]
\subfigure[~Single photon input $\ket{1}$]{
\includegraphics[width=115px]{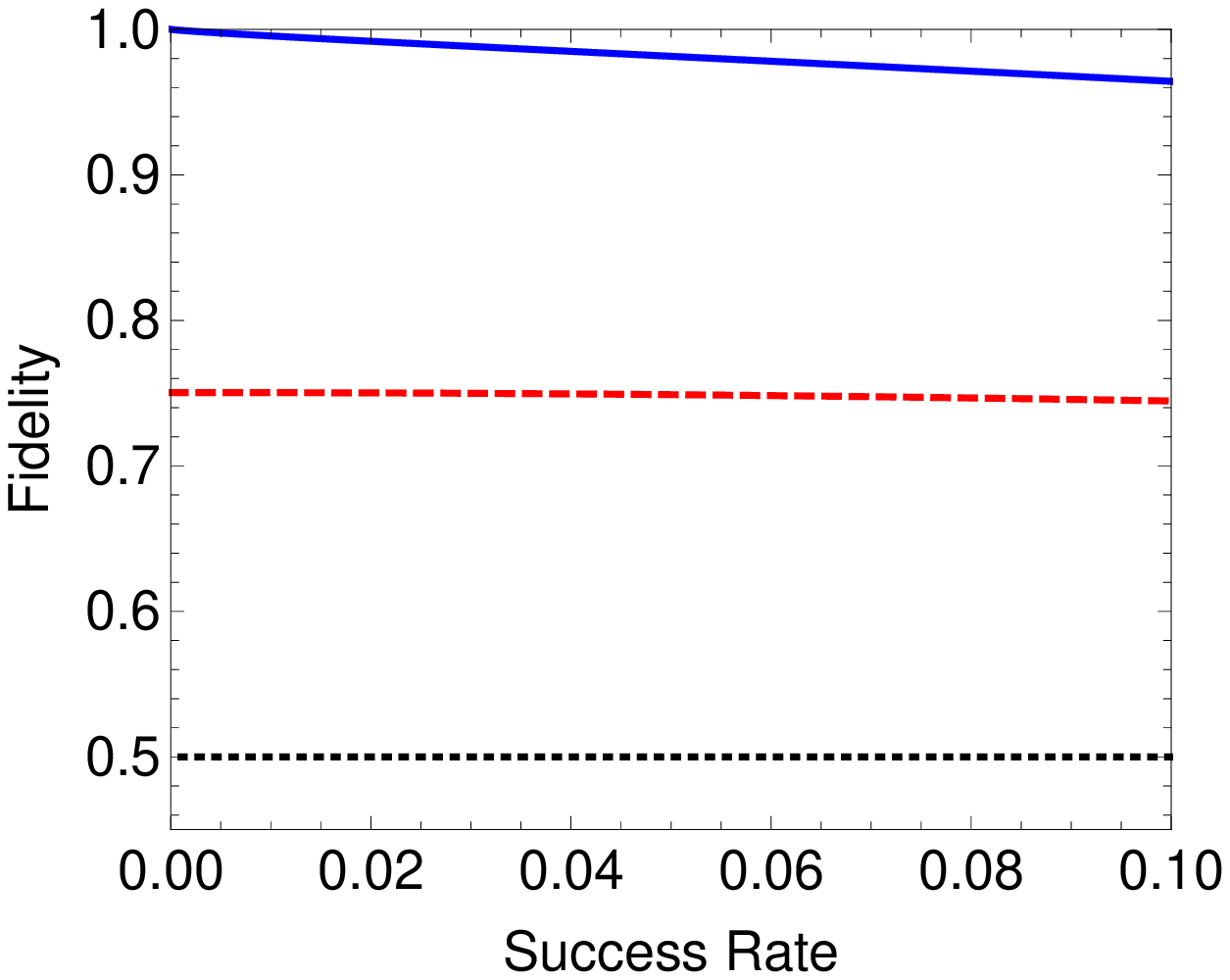}}
\subfigure[~Coherent state input $\ket{\beta=1}$]{
\includegraphics[width=115px]{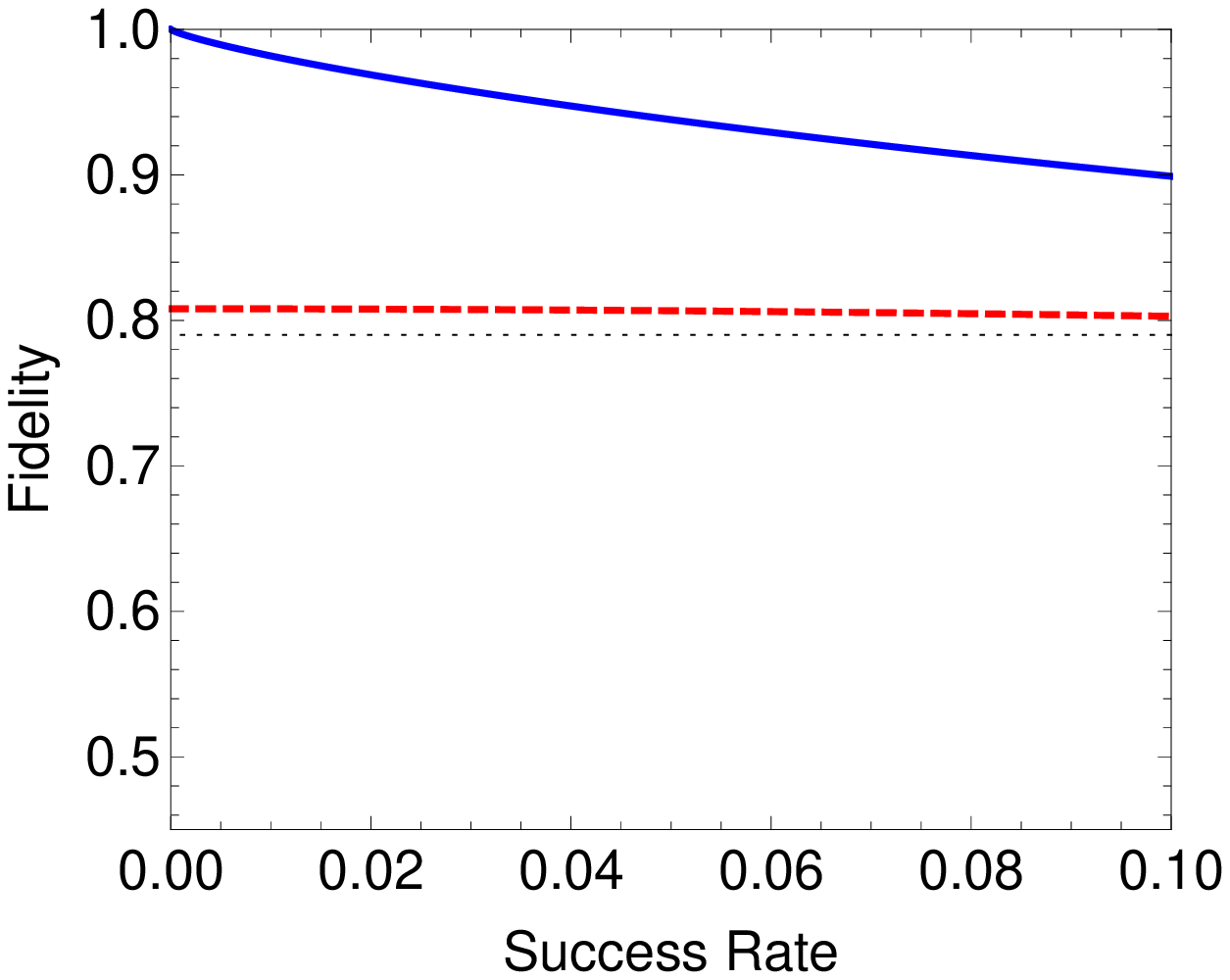}}
\caption{$F(\epsilon)$ vs $P(\epsilon)$ for (a) single photon input $\ket{1}$ and (b) coherent state input $\ket{\beta=1}$ for the operation $1+1.5\hat{a}-1.5\hat{a}^\dagger$ with (blue) a perfect single photon ancilla and (red, dashed) an imperfect single photon with $\eta=0.8$ for the homodyne measurement window $10^{-3}\le\epsilon\le1$. The setup is optimized for the largest $F$. Here no correction is considered. $F=0.95$ and $P=0.10$ is achieved for a perfect single photon input and $F=0.91$ and $P=0.10$ is achieved for a coherent state input. }
\label{fig:FvsP}
\end{figure}

So far we have been concerned in ideal projections onto quadrature eigenstates. This is just an idealization, and in practice such a projection onto a quadrature eigenstates has a zero probability of success. In practice it needs to be approximated by performing a homodyne detection and post-selecting upon detecting a value which falls closely into a small interval $\epsilon$ around the sharp target value $x_0$, which necessarily reduces the quality of the gate as a cost. The fidelity with the target state $\ket{\psi_t}$ of this realistic gate applied to state $\rho_\mathrm{IN}$   can be expressed as $F(\epsilon)=\int_{x_0-\epsilon}^{x_0+\epsilon}\mathrm{d} x\mathrm{Tr}[(\ket{\psi_t}\bra{\psi_t}\otimes\ket{x}_L\bra{x})U_\mathrm{BS} \rho_\mathrm{IN}\otimes \ket{1}_L\bra{1} U_\mathrm{BS}^\dagger ]/P(\epsilon)$, where the probability of success is  $P(\epsilon)=\int_{x_0-\epsilon}^{x_0+\epsilon}\mathrm{d} x \mathrm{Tr}[_L\bra{x}U_\mathrm{BS}\rho_\mathrm{IN}\otimes \ket{1}_L\bra{1} U_\mathrm{BS}^\dagger \ket{x}_L]$. In Fig.~\ref{fig:FvsP}, the fidelity and the probability of success of the operation $1+1.5 \hat{a}-1.5\hat{a}^\dagger$ applied to a single photon and to a coherent state are plotted both for a perfect ancilla $\eta=1$ and a realistic ancilla $\eta=0.8$. We can see that although there is a visible drop of fidelity for a perfect single photon ancilla when we increase $\epsilon$, the fidelity still remains quite high and obviously above the classical threshold. Furthermore, the reduction of fidelity is less prominent for the imperfect ancilla, which is very promising for the eventual experimental implementation.

Our scheme can be compared to the previous one proposed in \cite{SLeePRA2010}, which
employs inline coupling into a parametric downconverter, interferometer
and two single photon detectors. Apart from the feasibility, our scheme can exhibit success rates of around $0.05$, while the previous
proposal did not surpass $10^{-12}$, mainly due to low rate of the
down-conversion process.

\section{Multiple gates for applications}
\subsection{Conditional generation of cubic non-linearity}
As a prominent example, a  non-Gaussian qubic Hamiltonian up to the quadratic expansion can be achieved as:
\begin{eqnarray}
&\exp[i \chi \hat{X}^3]\approx 1+i \chi \hat{X}^3-\frac{\chi^2}{2}\hat{X}^6\propto\nonumber\\
&(1-(\tfrac{\chi}{-1+i})^{1/3}\hat{X})(1+(\tfrac{\chi}{1-i})^{1/3}\hat{X})\nonumber\\
&(1-(-1)^{-2/3}(\tfrac{\chi}{-1+i})^{1/3}\hat{X})
(1-(\tfrac{\chi}{1+i})^{1/3}\hat{X})\nonumber\\
&(1+(\tfrac{\chi}{-1-i})^{1/3}\hat{X})(1-(-1)^{-2/3}(\tfrac{\chi}{1+i})^{1/3}\hat{X})
\end{eqnarray}
where $\chi$ is the nonlinearity strength, and the attenuation is omitted for simplicity. This  second-order expansion is sufficient to achieve the  qubic nonlinearity for a general purpose~\cite{Marek2011}. {Exploiting the emerging single photon guns, it will be the first step towards controlled nonlinear dynamics of quantum oscillator. The identification of hidden non-classical features of quantum states produced by the cubic nonlinearity has been proposed \cite{Yukawa2013}.}

\subsection{Arbitrary wave-function generation}
It is well known that any quantum state can be approximated with an arbitrarily high precision by a finite superposition of fock states up to $N$-th order as $|\psi\rangle = \sum_{n=0}^{N}c_n \hat{a}^{\dagger n}/\sqrt{n!}\ket{0} $. We observe that this state can be constructed by a polynomial of $\hat{a}^\dagger$ applied to the vacuum state \cite{Dakna1998}. This operation is achieved by the repeated application of the elementary operation $1+\lambda \hat{a}^\dagger$, which is a special case of Eq.~(\ref{eq:aftercorrection})  with $B=0$. Complementary to this approach, we can also use the continuous-variable operators to build not the discrete Fock state expansion of the state but rather the continuous-variable wave function of the state. The wave function of general state in the coordinate representation can be simply expressed as:
\begin{equation}\label{}
    \psi(x) =\langle x|\psi\rangle= \sum_{n=0}^{N} \frac{c_n H_n(x)}{\pi^{1/4}\sqrt{2^n n!}}e^{-x^2/2}\equiv G(x)\braket{x}{0},
\end{equation}
where $G(x)=\sum_{n=0}^{N} \frac{c_n H_n(x)}{\sqrt{2^n n!}}$ and $H_n(x)$ are Hermite polynomials. Therefore we can write $\ket{\psi}=G(\hat{X})\ket{0}$.
This is simply a wave function of the vacuum state multiplied by a $N$th order polynomial of $x$, which is exactly obtained by $N$-fold application of the $X$ gates. The number of required operations can be reduced by attempting to generate a suitably squeezed version of the target state and then manipulating the Gaussian envelope by another squeezing operation \cite{Manzies2009}. Therefore, the $X$ gate can be seen as a universal elementary gate sufficient for general state preparation - the continuous counterpart of the particle-like single photon addition.

\begin{figure}[tp]
\includegraphics[width=240px]{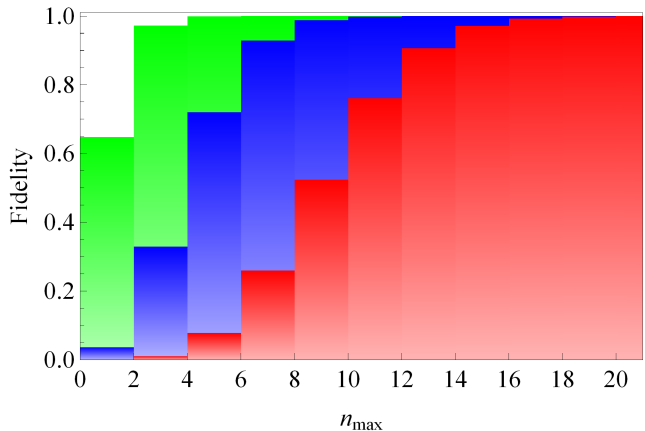}
\caption{Fidelity of the ideal coherent state superposition $N_c(\ket{\beta}+\ket{-\beta})$ with the generated cat states. Depending on the highest photon number $n_\mathrm{max}$ which coincides the repetition number of X-gate, we can achieve the approximate cat state very efficiently. Green, blue and red bars correspond to $\beta=1,2,3$, respectively.}
\label{fig:cat}
\end{figure}
To demonstrate the state generation aspect of our scheme, let us attempt to generate an equal superposition of coherent states, $N_c(\ket{\beta}+\ket{-\beta})$, where $N_c=(2+2e^{-2\beta^2})^{-1/2}$ is the normalization factor. This quantum state is an important resource in quantum information processing and fundamental tests of quantum mechanics~\cite{CochranePRA1999, JeongPRA2002, RalphPRA2003, GerryPRA1995}, and has been realized experimentally for $\beta\lesssim 2$~\cite{YurkePRL1986,OurjoumtsevNature2007,TakahashiPRL2008,OurjoumtsevNatP2009}. It alternatively can be written as $N_c e^{-\beta^2/2}(\exp[\beta \hat{a}^\dagger]+\exp[-\beta \hat{a}^\dagger])\ket{0}=N_c'(n_\mathrm{max})\sum_{n=\mathrm{even}}^{n_\mathrm{max}}2(\beta\hat{a}^\dagger)^n/n!\ket{0}$, where $N_c'(n_\mathrm{max})$ is a normalization factor for a finite expansion up to the maximum photon number $n_\mathrm{max}$ in a truncated form. This state is generated by the following polynomial of $\hat{a}^\dagger$ on the vacuum state; $\sum_{n=\mathrm{even}}^{n_\mathrm{max}}2(\beta\hat{a}^\dagger)^n/n!$. The dependence of the fidelities on $n_\mathrm{max}$ with the exact even cat state are drawn in Fig.~\ref{fig:cat}. We note that for $n_\mathrm{max}=16$, we can achieve the fidelity of $0.993$ for $\beta=3$. An odd cat state can be constructed in a completely equivalent way. We also note that no attenuation effect exists in the state generation due to the initial vacuum state the scheme acts on.

\subsection{Multiple X-gates in a single shot operation}
Implementing a potential $F(\hat{x})$ by the sequential application of X-gates is accompanied by an exponential decrease of the probability of success. This issue can be overcome by applying the total potential consisting of several X-gates directly in a single step. First, a specific ancillary state $f(\hat{X}_L)\ket{0}_L$, where $f(\hat{x}) = F(-\hat{x}/\kappa)$, can be generated off-line using X-gates, similarly as in \cite{Marek2011}. After a QND coupling between the ancilla and the oscillator, the ancillary mode is measured by homodyne detection and the target operation is achieved:
\begin{align}
&_L\bra{x_0=0}U_\mathrm{QND}f(\hat{X}_L)\ket{0}_L=_L\bra{x_0=0}f(-\kappa \hat{X})U_\mathrm{QND}\ket{0}\nonumber\\
&=f(-\kappa \hat{X})_L\bra{x_0=0}U_\mathrm{QND}\ket{0}=F(\hat{X})\exp[-\frac{1}{2}\kappa^2 \hat{X}^2].
\end{align}
The factor $\exp[-\kappa^2\hat{X}^2/2]$ can be compensated by a suitable squeezing of the ancilla as before. The same approach can be applied to the operations based on the beam splitter interaction. In this scheme the unavoidable attenuation is suppressed as a side benefit. 


\section{Conclusions}
 We have presented a methodology for the conditional induction of various nonlinear potentials on quantum oscillators and conditional preparation of wave functions of the quantum oscillators. This method is based on the sequential application of the elementary X-gates supplied by the single-photon guns. Based on a wide class of emerging single photon guns \cite{Kuhn2002,Legero2004,Chen2006,Hijlkema2007,Bozyigit2011,Fortsch2013,Collins2013,He2013},
it is broadly applicable for various quantum oscillators (optical, atomic, or mechanical. The presented operation will therefore open a broad area of very anticipated investigation of controllable nonlinear dynamics of quantum oscillators.

\acknowledgments
R.F. and P.M. acknowledge a financial support from grant No. GA14-36681G
of Czech Science Foundation. K.P. acknowledges financing by the European Social Fund and the state budget of the Czech Republic, POST-UP NO CZ.1.07/2.3.00/30.0004.


\end{document}